\documentclass[conference]{IEEEtran}
\IEEEoverridecommandlockouts
\usepackage[normalem]{ulem}

\usepackage{booktabs} 
\usepackage{mathptmx} 

\newcommand{\ignore}[1]{}
\usepackage{fancyhdr}
\usepackage{hyperref}
\usepackage{multirow}
\usepackage{graphicx}
\usepackage{caption}
\usepackage{pifont}
\usepackage{amsmath}
\usepackage{float}
\usepackage{color}
\usepackage{array}
\usepackage[flushleft]{threeparttable}
\usepackage[bottom]{footmisc}

\usepackage{listings}

\newcolumntype{L}[1]{>{\raggedright\let\newline\\\arraybackslash\hspace{0pt}}m{#1}}
\newcolumntype{C}[1]{>{\centering\let\newline\\\arraybackslash\hspace{0pt}}m{#1}}
\newcolumntype{R}[1]{>{\raggedleft\let\newline\\\arraybackslash\hspace{0pt}}m{#1}}%

\hypersetup{draft}
\newcommand{\bheading}[1]{{\vspace{2pt}\noindent{\textbf{#1}}\hspace{2pt}}}

\makeatletter
\newcommand*\titleheader[1]{\gdef\@titleheader{#1}}
\AtBeginDocument{%
  \let\st@red@title\@title%
  \def\@title{%
    \bgroup\normalfont\large\centering\@titleheader\par\egroup
    \vskip1.5em\st@red@title}
}
\makeatother

\title{Power-Grid Controller Anomaly Detection with Enhanced Temporal Deep Learning}
\titleheader{The 18th IEEE International Conference on Trust, Security and Privacy in Computing and Communications (TrustCom'19)}

\begin{document}

\author{\IEEEauthorblockN{Zecheng He}
\IEEEauthorblockA{\textit{Princeton University} \\
Princeton, NJ \\
zechengh@princeton.edu}
\and
\IEEEauthorblockN{Aswin Raghavan}
\IEEEauthorblockA{\textit{SRI International} \\
Princeton, NJ \\
aswin.raghavan@sri.com}
\and
\IEEEauthorblockN{Guangyuan Hu}
\IEEEauthorblockA{\textit{Princeton University} \\
Princeton, NJ \\
gh9@princeton.edu}
\and
\IEEEauthorblockN{Sek Chai}
\IEEEauthorblockA{\textit{SRI International} \\
Princeton, NJ \\
sek.chai@sri.com}
\and
\IEEEauthorblockN{Ruby Lee}
\IEEEauthorblockA{\textit{Princeton University} \\
Princeton, NJ \\
rblee@princeton.edu}
}

\date{}
\maketitle



\begin{abstract}
Controllers of security-critical cyber-physical systems, like the power grid, are a very important class of computer systems. Attacks against the control code of a power-grid system, especially zero-day attacks, can be catastrophic. Earlier detection of the anomalies can prevent further damage. However, detecting zero-day attacks is extremely challenging because they have no known code and have unknown behavior. Furthermore, if data collected from the controller is transferred to a server through networks for analysis and detection of anomalous behavior, this creates a very large attack surface and also delays detection.

In order to address this problem, we propose Reconstruction Error Distribution (RED) of Hardware Performance Counters (HPCs), and a data-driven defense system based on it. Specifically, we first train a temporal deep learning model, using only normal HPC readings from legitimate processes that run daily in these power-grid systems, to model the normal behavior of the power-grid controller. Then, we run this model using real-time data from commonly available HPCs. We use the proposed RED to enhance the temporal deep learning detection of anomalous behavior, by estimating distribution deviations from the normal behavior with an effective statistical test. Experimental results on a real power-grid controller show that we can detect anomalous behavior with high accuracy ($>$99.9\%), nearly zero false positives and short ($<$360ms) latency.

\end{abstract}

\section{Introduction}\label{sec:intro}

The power-grid is a critical infrastructure, and attacks on it can cripple society, affecting national security, economic competitiveness and societal interactions. In 2015, Ukraine experienced cyber attacks against its power-grid system \cite{2015powergridattack}. During this attack, 30 substations were switched off and 230 thousand people were left without electricity. The U.S. Government Accountability Office (GAO) questioned the current adequacy of defenses against power-grid attacks, and the North American Electric Reliability Corporation (NERC) has recognized these concerns \cite{sridhar2012cyber}. With power-grid substations controllable through cyberspace transactions, the concern about cyber attacks on the physical power-grid systems is a real and serious threat, which is the focus of our paper.

There have been detection approaches proposed to protect the power-grid system by examining the physical sensors \cite{manandhar2014detection, valenzuela2013real}. However, the approaches examining sensors only consider the physical part but ignored the cyber attacker. Furthermore, they focused on the algorithms but neglected the system design. Previous work on the cyber-security of power-grid systems focused on protecting the networks and data transmission \cite{wang2011security, sikdar2011defending}. Each of these solved a specific loophole of secure communication in the power-grid system, but did not protect controllers and cannot detect new attacks.


Besides the physical sensors and network, protecting the control code running on the industrial programmable logic controllers (PLCs) is an essential and fundamental task in protecting the power-grid systems. No matter how the adversary spreads the malware or bad programs through system vulnerabilities, his ultimate goal is taking control of the power-grid infrastructures, destroying them or turning them off, and thus inducing huge physical impact. This attack strategy on the industrial controller has been shown in the Stuxnet \cite{langner2011stuxnet} and BlackEnergy \cite{2015powergridattack} attacks. Therefore, rather than physical attacks on the physical power-grid system, we consider cyber attacks that hijack the controller code that controls the actuators of the physical system.

The problem of protecting the controllers in the power-grid system is a critical unsolved problem. It is challenging because \ding{172} there is no prior-knowledge about the attacks, especially in "zero-day" attacks. \ding{173} Various controllers, e.g. ABB, Siemens and Wago, and OS, e.g. Windows, Unix and proprietary OS (SIMATIC WinCC), are widely used in modern power-grid systems, exposing a large attack surface. \ding{174} The power-grid systems usually implement weak anti-virus and integrity checking mechanisms on both executed code and transferred data, due to the computation capacity and hard-realtime constraints \cite{humayed2017cyber}. \ding{175} Many concurrent threads are running on the controller simultaneously. Stealthy attack programs that have infected the controller can hide within these threads. 

In this paper, we propose a new method, Reconstruction Error Distribution (RED) of Hardware Performance Counters (HPCs), to detect anomalies in power-grid controllers. Our proposed defense provides comprehensive and reliable protection against unknown attacks. It does not look for specific attacks or triggers, but rather takes a holistic view of controller operation. Specifically, our proposed approach first automatically learns the normal behavior of the controller, and generates a corresponding behavioral model and a RED profile of the controller. Then we use the generated model, enhanced by an effective statistical test, to detect the RED deviation from the controller's normal behavior.

Our main contributions in this paper are:

\begin{itemize}

\item We propose Reconstruction Error Distribution (RED) of Hardware Performance Counters (HPCs) as a profile of the controller's normal behavior, and show it is effective in detecting anomalies in power-grid controllers.

\item We implement a new data-driven, learning-based, statistically-enhanced system based on RED for detecting anomalies in the power-grid controllers. No prior-knowledge of attacks is needed in either the training or inference, thus it can be used to detect zero-day attacks.



\item We evaluate our proposed defense on real power-grid controllers in multiple dimensions. We simulate publicly reported attack functionalities \cite{langner2011stuxnet, case2016analysis} and other attacks, and show the superiority of our proposed solution, compared to other machine learning approaches.

\end{itemize}

In Section \ref{sec:background}, we provide background material. In Section \ref{sec:SecurityConsiderations}, we articulate our threat model. In Section \ref{sec:Method}, we present our approach in detail. In Section \ref{sec:Evaluation}, we present the accuracy and performance evaluation, respectively. We discuss related work in Section \ref{sec:relatedwork} and conclude in Section \ref{sec:conclusion}.

\section{Background}\label{sec:background}



\subsection{Power-Grid System Architecture and Vulnerability}

\begin{figure}[h]
 \centering
  \includegraphics[width=0.8\linewidth]{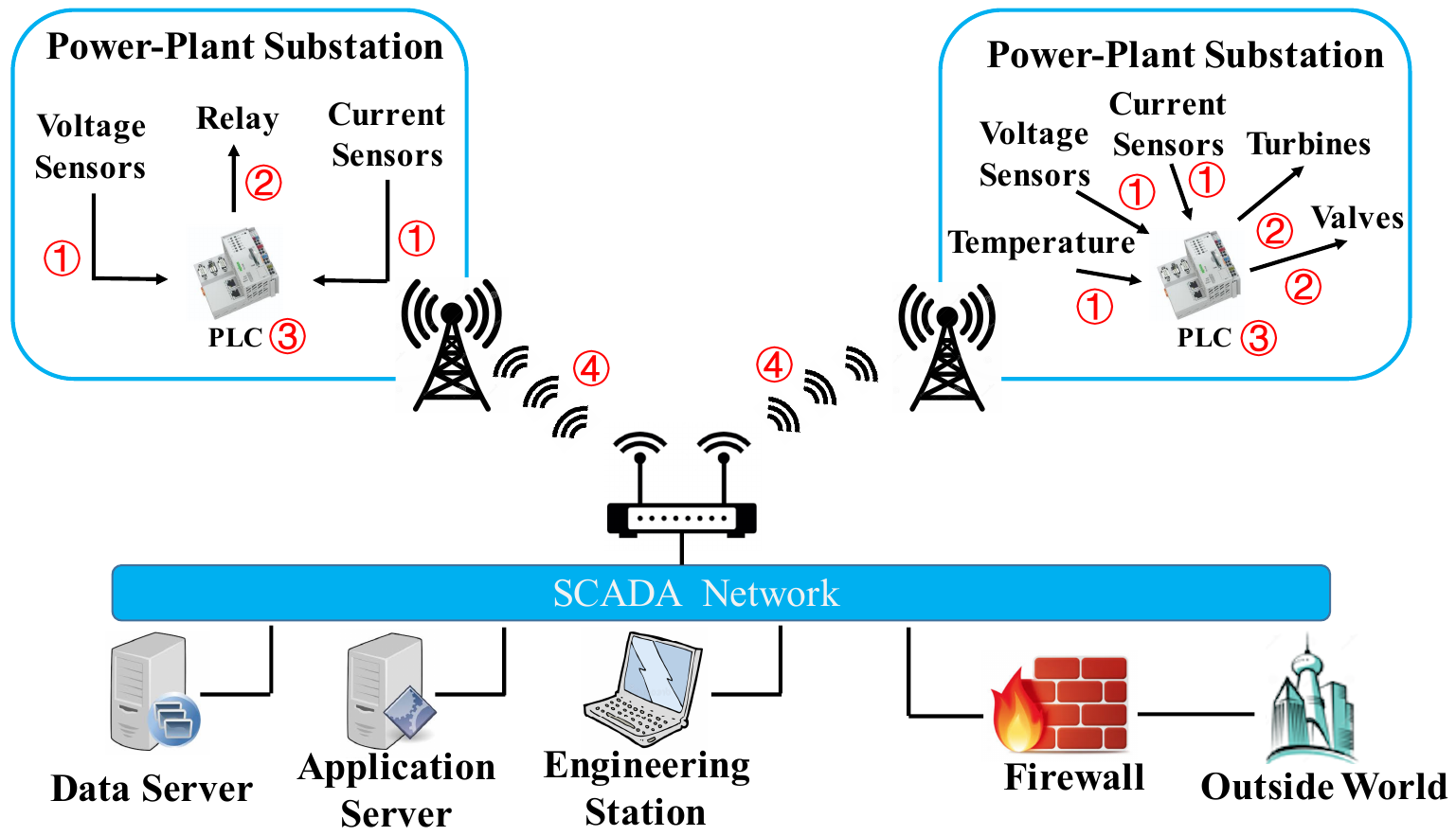}
  \caption{A diagram of a power-grid system. Attacks can happen in \ding{172} sensors \ding{173} actuators \ding{174} controllers and \ding{175} networks.}
  \label{fig:SCADA}
\end{figure}

Figure \ref{fig:SCADA} shows a SCADA (Supervisory Control And Data Acquisition) system controlling a power-grid \cite{SCADA}. While the rest of the SCADA network and components are similar to general information processing systems, the power-plant substations are the cyber-physical systems we focus on. The key components in these substations are physical sensors (e.g. voltage, current), actuators (e.g. relays, valves) and programmable logic controllers (PLCs). The PLC obtains inputs from the sensors, and outputs control signals to the actuators according to its control logic or control code. The substations are connected through local networks.




Attacks against the power-grid system can happen in sensors, actuators, controllers and the interconnected networks illustrated in Figure \ref{fig:SCADA}. Attacks against the physical actuators or sensors usually require physical accesses to the devices, and are not in the scope of this paper. The ultimate goal of the network attacks against the power-grid is launching the malware, compromising the controller to induce physical damage. Hence, we focus on the detection of the controller's abnormal behavior on the power-grid.

\subsection{Programmable Logic Controller (PLC)}\label{sec:PLC}
\begin{figure}[h]
 \centering
  \includegraphics[width=0.8\linewidth]{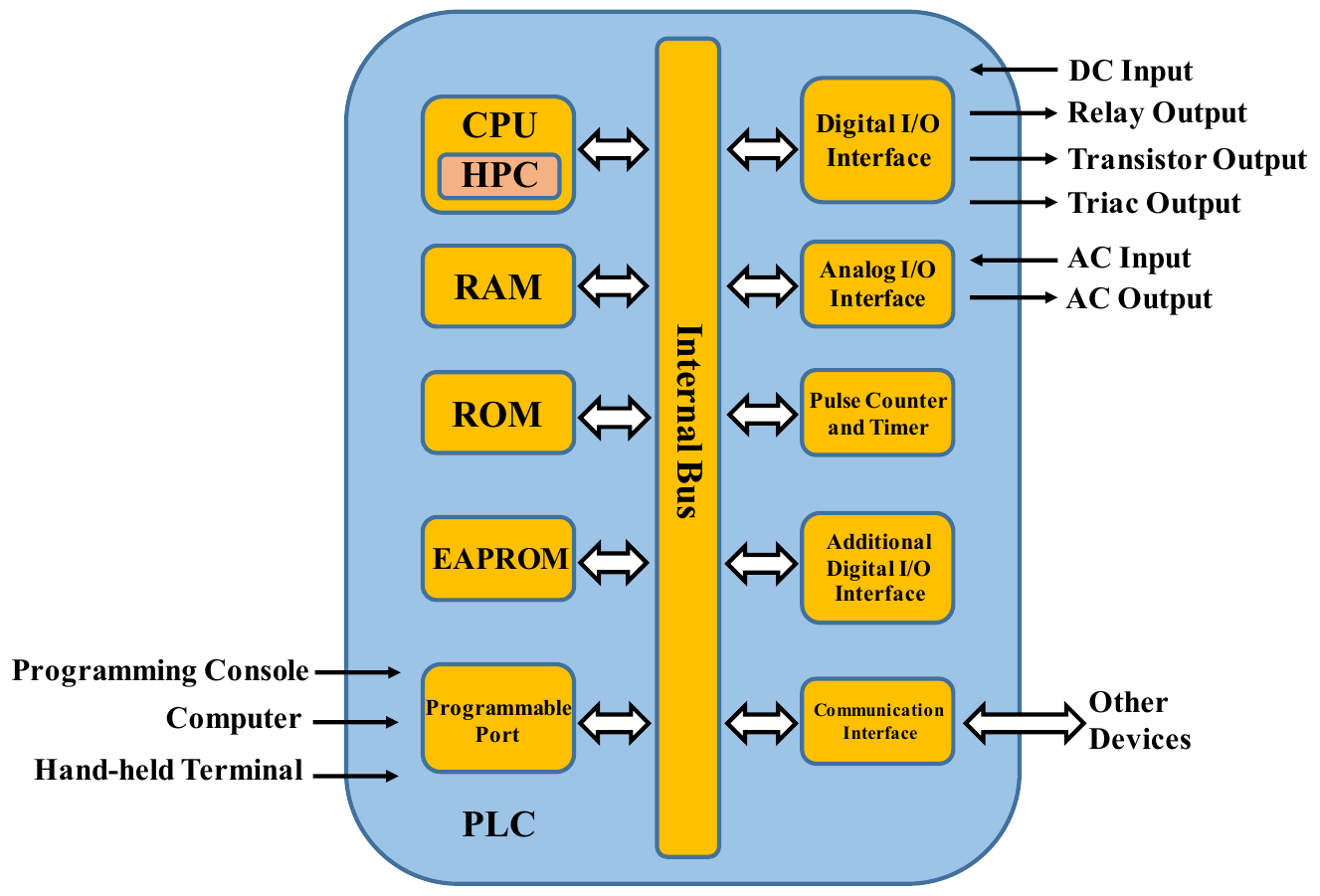}
  \caption{Programmable logic controller block diagram \cite{PLCdiag}}
  \label{fig:PLCdiag}
\end{figure}

A programmable logic controller (PLC) is a digital device, equipped with microprocessors, which is widely used as the controller in a power-grid subsystem. Figure \ref{fig:PLCdiag} shows the block diagram of a PLC \cite{PLCdiag}. A PLC is composed of several subsystems, e.g. CPU, internal memories and communication interface. A PLC can take physical sensor (and other) inputs from the target industrial devices, make decisions and send commands to control a large range of devices and actuators.

The PLC used in our experiments is a Wago PLC with an ARM Cortex A8 processor. A custom real-time Linux OS runs on this PLC. This PLC is widely used as the controller in a power-grid system.

\subsection{Hardware Performance Counters}\label{sec:HPCbackground}

Controller hardware events, such as the number of executed instructions, cache operations and misses, are automatically and securely counted and stored in Hardware Performance Counters (HPC) -- a set of special-purpose registers. HPCs are now widely available in commodity processors, e.g., Intel, ARM and PowerPC.

In this work, we use low-level hardware events to monitor the controller behavior. Hardware performance counters provide a high tamper-resistant and low-cost way to monitor the controller behavior. First, unlike anti-virus software, HPCs are hardware that cannot be directly accessed and modified by malware \cite{wang2016hardware}. This characteristic of HPCs significantly increases the tamper-resistant property. Second, HPCs automatically record the hardware events, requiring no extra logging or recording. Reading from HPCs is more efficient than hashing the whole control code, making it low-cost with negligible overhead.


\section{Threat Model}\label{sec:SecurityConsiderations}

We consider cyber attacks that hijack the controller code that controls the actuators of the physical system, making the controller behave abnormally. We do not explicitly consider the attacks against networks in power-grid systems, because the ultimate goal of the network attacks is corrupting the controller, which has been shown in the Stuxnet \cite{langner2011stuxnet} and BlackEnergy \cite{2015powergridattack} attacks. Thus these attacks can be detected if we can detect the controller's anomalous behavior. We highlight the key points of our threat model in detail:


\bheading{Our threat model specifically includes zero-day attacks.} We assume that there is no prior-knowledge of attack code, and the way the adversary hijacks the controller. We only assume the goal of the attacker is to hijack the controller and run unauthorized code on the hijacked controller. Note that we do not make any assumptions about the type of unauthorized code, i.e., it can be malicious control logic, worms, trojans or spyware, etc. Therefore, though not explicitly targeted, availability and confidentiality breaches are also included in our threat model. Furthermore, we assume the normal behavior of the controller can be collected in a clean environment and used for training a deep learning model.

\bheading{Attacker Capabilities.} We consider an active attacker who can hijack the controller code, add to, delete or modify the programs running on the controller. Consequences of the attacks include controller failures, incorrect outputs to the controlled physical actuators and other subsystems. The attacker is able to bypass the firewall through system vulnerabilities or client-side attacks, e.g the misuse of unauthorized USB drive in Stuxnet. The attacker can either access the victim system remotely (e.g. via network or Botnet) or physically (e.g. USB). The attack code can be polymorphic and stealthy.

\bheading{Target Systems.} We consider a power-grid system in our threat model, especially the cyber-physical power substation with a PLC. We assume the power-grid substation implements weak anti-virus and integrity checking mechanisms on both executed code and transferred data, due to the computation capacity and hard-realtime constraints \cite{humayed2017cyber}. We assume the low-level hardware performance counters are accessible and trusted. This assumption is reasonable because hardware registers are harder to tamper with than software. The target system is relatively stable, i.e. the code running on the controller changes infrequently. This assumption is reasonable because the control processes, once downloaded to the controller, usually are not updated very often \cite{lee2015past}.







\section{Detection Methodology}\label{sec:Method}





\subsection{Overview}
\label{sec:solution:overview}

\begin{figure}[h]
 \centering
  \includegraphics[width=0.8\linewidth]{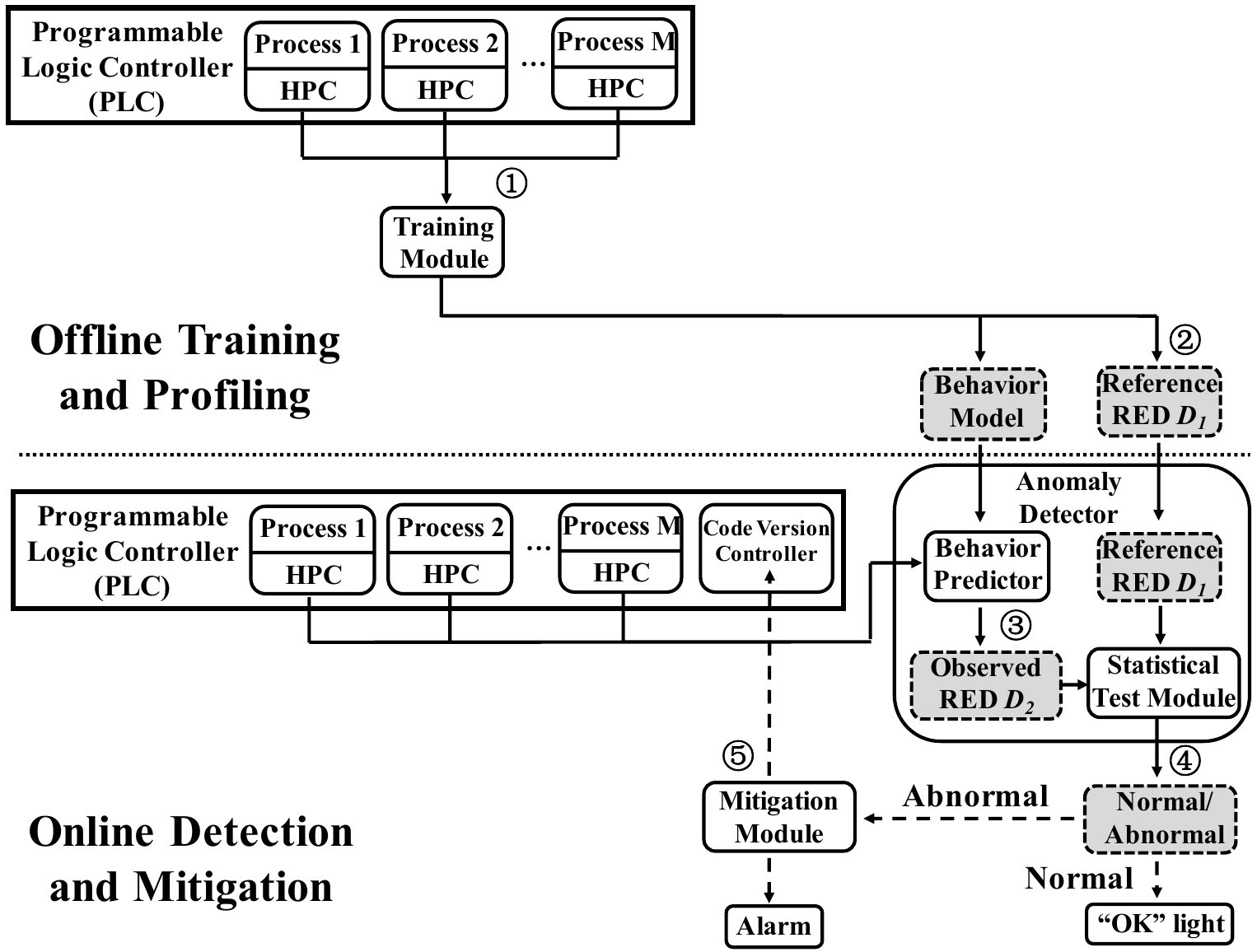}
  \caption{Overview of our proposed deep temporal model and statistical test based method. Offline profiling: \ding{172} Train a deep learning model with only normal data, to predict the normal behavior of the controller. \ding{173} Collect the reference reconstruction error distribution (RED) $D_1$ of the baseline behavior. Online detecting: \ding{174} Monitor the real behavior of the controller and compute the observed reconstruction error distribution (RED) $D_2$. \ding{175} Statistically test if $D_1$ and $D_2$ are the same distribution. If not, trigger the \ding{176} response and attack mitigation module.}
  \label{fig:architecture}
\end{figure}

We consider the attack scenario in which the attacker breaches the integrity of the control code and causes the controller to function incorrectly in the power-grid system. Our key idea is, the normal behavior of the controller in a power-grid system is predictable using a temporal deep learning model and low-cost HPC features. We define \textbf{RED}, \textbf{Reconstruction Error Distribution}, of hardware performance counters, as a robust way to detect controller anomalies. The deviation of the real monitored behavior from the predicted behavior indicates the anomalies. We further use a statistical test to emphasize such deviations. We show our proposed architecture in Figure \ref{fig:architecture}.


There are two phases, i.e. \textit{offline training and profiling} and \textit{online detection and mitigation}, and five steps in the power-grid controller anomaly detection. The offline phase consists of two main steps:

\begin{enumerate}
\item Training a deep learning sequence predictor to predict future normal controller behavior. The sequence predictors we explore are Long Short-Term Memory (LSTM) and Conditional Restricted Boltzmann Machine (CRBM).
\item Calculating the baseline RED $D_1$ as the reference distribution. In our experiments, we use the squared error of the predicted behavior $v_1$ and the observed behavior $v_2$, i.e. $|v_1-v_2|^2$, as the reconstruction error.
\end{enumerate}
The online detection phase consists of three main steps:
\begin{enumerate}
\item Using the sequence predictor online to predict the future behavior using the historical behavior. Calculating the RED $D_2$ of the observed (testing) behavior which can be normal or abnormal.
\item Applying the statistical test on $D_1$ and $D_2$ to determine if they are from the same distribution.
\item If an anomaly is detected, the anomaly response module is triggered. One action is to switch to a "safe but not updated" version of the control code, and send out an alarm.
\end{enumerate}


Next, we highlight some challenges and concerns in designing an effective protection for power-grid controllers.

\bheading{Availability of only normal data.} Since attack data is very sensitive, it is very hard to obtain. Furthermore, zero-day attacks have no attack code nor known behavior. To address this challenge, our proposed approach only needs normal data. Specifically, it predicts the future behavior of the controller through a deep temporal model, and investigates the deviation between the prediction and observation error distributions.


\bheading{Capturing high-level controller behavior with low-level HPC features.} Although using HPC measurements benefits from the low-cost and high tamper-resistant properties, it is challenging to appropriately capture and characterize high-level controller behavior from the low-level hardware events. Conventional machine learning or clustering based anomaly detection approaches require carefully designed heuristic features, however, high-level semantic features are typically hard to handcraft from low-level measurements.

\bheading{Comprehensive protection.} To provide comprehensive protection, we monitor all threads running on the controller. HPCs automatically record the hardware events, requiring no extra logging or recording. Reading from HPCs is more efficient than hashing the whole control code, making it low-cost with negligible overhead to monitor all running processes. There are fewer threads (∼10-100) running on the power-grid controller than a general- purpose system, making it possible to monitor all threads.

\bheading{Polymorphic and stealthy attacks.} To bypass any static malware analysis, the adversary can write code variants that are functionally equivalent (polymorphic codes), or hide by prolonging the time-span, or mimicking the normal behavior. Our detection algorithm must characterize the program by its functionality and behavior.

\subsection{Offline Training and Reconstruction Error Profiling}
\label{sec:solution:offline}


\bheading{Model selection.} Among the deep learning models, Recurrent Neural Network (RNN) and its variation, Long Short-Term Memory (LSTM), \cite{hochreiter1997long} have become the most powerful ones in modeling sequential data. An LSTM cell has three gates that control information flow: the forget gate, the input gate and the output gate. The forget gate controls the amount of previous information remaining in the cell memory. The input gate controls the amount of new information flowing to the cell memory, and the output gate controls the amount of information flowing to the output. Thus, LSTM automatically determines what information to "remember" and "forget".

We also explore the Conditional Restricted Boltzmann Machine (CRBM). RBM is a shallow, two-layer stochastic neural network. Each node in the visible layer takes a low-level feature (e.g. HPC values), each node in the invisible layer represents a high-level learned feature (e.g. controller behavior). However, the vanilla RBM does not capture the temporal information, therefore in this work, we use a variant of RBM which considers historical information, i.e., Conditional-RBM (CRBM). We compare LSTM and CRBM used as the controller behavior predictors in our evaluation.

\bheading{Model training.} A temporal deep learning model is trained, illustrated as \ding{172} in Figure \ref{fig:architecture}, to predict the controller's future behavior based on its historical behavior.

In the training, we first collect a set of sequences of the controller behavior measurement $\{S_{i}\}_{i=1}^N$, which are HPC measurements in our scenario, in a clean environment. $N$ is the number of total sequences. Each sequence $S_i$ consists of $T$ continuous behavior measurements, i.e. $S_{i}=[S_i^1,...,\ S_i^{T}]$. In our experiments, each behavior measurement $S_i^t$ is a vector consisting of 4 HPC readings for 23 threads, i.e. a vector of dimension 4*23=92. At time $t$, the deep learning model predicts $S_i^{t+1}$ using behavior history $[S_i^1, ..., \ S_i^{t}]$. We denote this prediction as $P_i^{t+1}$. The loss function is defined as the accumulated prediction errors, i.e.

{\small
\begin{align}
loss=\frac{1}{N} \frac{1}{T-1} \sum_{i=1}^{N} \sum_{t=2}^{T} (S_i^{t}-P_i^{t})^2
\end{align}
}%


Intuitively, since $\{S_{i}\}_{i=1}^N$ are normal behavior collected in the clean environment, the loss penalizes the incorrect prediction of normal behavior. We train this model to minimize the loss function with Stochastic Gradient Descent (SGD).


\noindent\bheading{Reconstruction error distribution profiling.} After training the model, we profile the normal behavior in terms of reconstruction error distribution (RED), illustrated as \ding{173} in Figure \ref{fig:architecture}. First, a longer reference sequence of controller behavior measurement, $R=[R^1,..., R^{T'}]$, is collected in the clean environment. Note that $R$ is another sequence rather than the previous $S_{i}$, consisting of $T'$ time frames. Each time frame $R^i$ is a vector of dimension 4*23=92 in our experiment. Second, at time frame $t$, we use the trained model to predict time frame $t+1$,...,\ $t+L$ using the corresponding history behavior. We denote the prediction as $P^{t+1},...,P^{t+L}$ respectively. The reconstruction error is defined as:

\vspace{-0.5cm}
{\small
\begin{align}
E(t)= \sum_{i=t+1}^{t+L} (R^{i}-P^{i})^2 \label{equ:reconstructionerror}
\end{align}
}

We define the \textbf{distribution} of \{E(1), E(2), E(3)...\} as the \textbf{Reconstruction Error Distribution (RED)}. Using the RED as the profile has several advantages. \ding{172} It is more robust to the noise in the measurement, because noise makes a single prediction error vary significantly but the distribution remains stable. \ding{173} It simplifies the model learning process. Since we profile the error distributions, we no longer need to train a "perfect" predictor. \ding{174} It broadens the scope of target systems by decreasing the required level of predictability. By using RED, we allow imprecisions in the prediction, reducing the required predictability of the system to a mild level.

In the profiling, collections of the reference RED are required to be gathered in a clean environment. This can be done by collecting the HPC sequences immediately after a trusted version of the control code is uploaded. The gathered HPC sequences are sent to the trained model to calculate the reference RED. These REDs are then stored in secure memory, and used as the references in the online detection. To better represent the RED by the empirical samples, multiple reference REDs can be generated and compared with the observed RED in the online detection phase.

\bheading{Model evolution.} To accommodate legitimate controller behavior drift, e.g. an update of the control code, the deep learning model needs to be retrained or fine-tuned online. Note that either in the initial training or online fine-tuning, the deep learning model must be trained with HPC measurements from a clean (no attack) environment. Otherwise, the trained model cannot predict the normal behavior correctly.

\subsection{Online Hijack Detection and Mitigation}
\label{sec:solution:online}

The online hijacking detection is illustrated as steps \ding{174} and \ding{175} in Figure \ref{fig:architecture}. The controller behavior, in terms of HPCs, is dynamically monitored at runtime. Similar to the offline profiling phase, the runtime gathered HPC sequences are sent to the deep learning module for prediction. The same form (Eq. \ref{equ:reconstructionerror}) of reconstruction errors are calculated and sent to the Kolmogorov-Smirnov (KS) test module, along with the reference reconstruction errors gathered in the offline phase. All the computations are performed outside the controller, on the attached trusted anomaly detector module, because if the controller is hijacked, the results computed by it cannot be trusted.



Our detection system can be integrated with different mitigation approaches. For example, after a controller hijacking attack is detected, an alarm is sent out. Another mitigation response is switching to a "safe version" of controller code in a ROM, which may not be up-to-date but is free from attack. Other responses are possible, such as controlling the actuator settings, while checking that an attack has actually occurred rather than a false alarm.

\subsection{Anomaly Detection through RED}
\label{sec:solution:KS}

We now address the problem: how to effectively determine anomalies from the reconstruction error distribution (RED)?

\begin{figure}[h]
\centering
  \includegraphics[width=\linewidth]{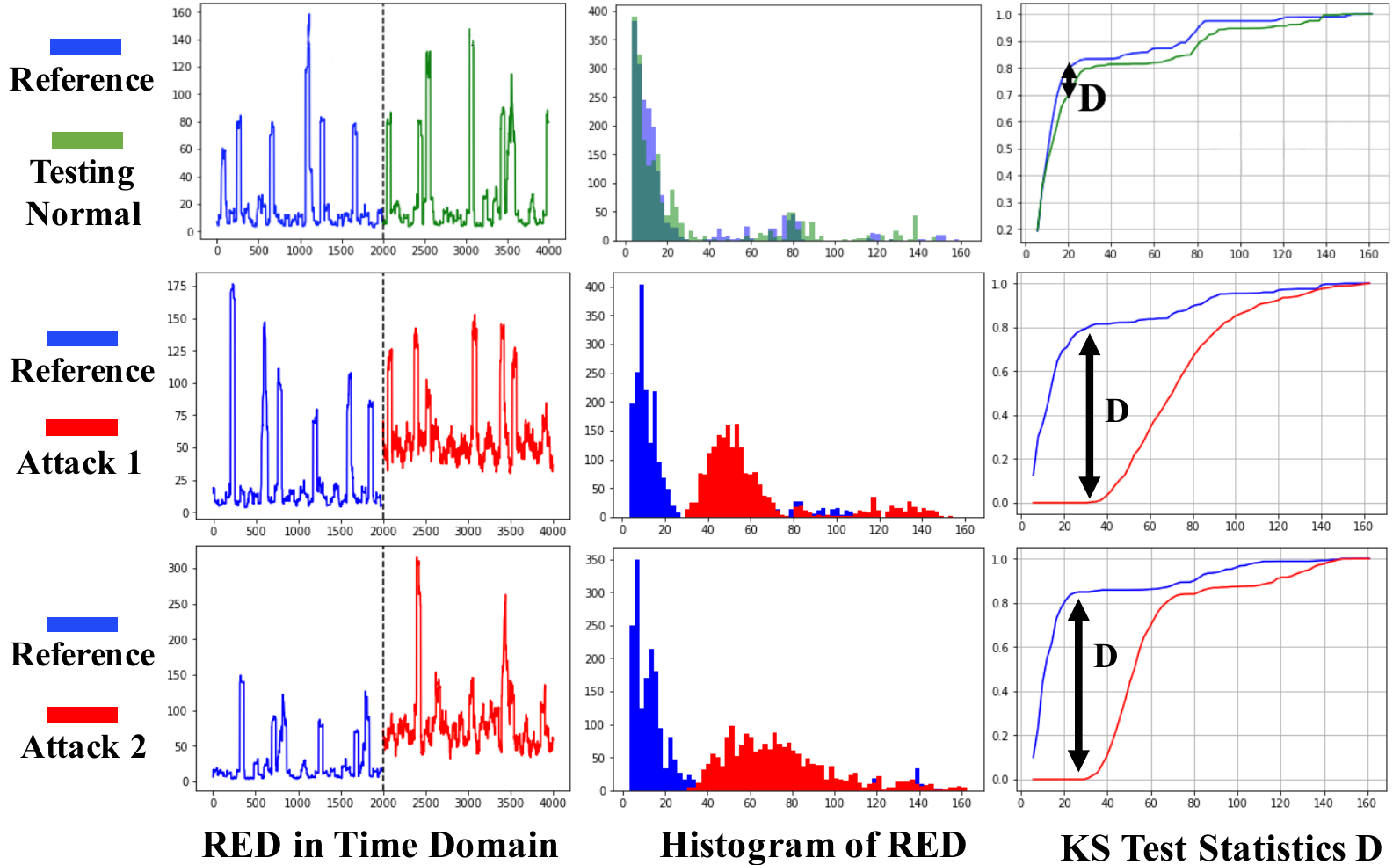}
  \caption{Effectiveness in detecting attacks. Left: REDs in the \textit{time domain}. Middle: Histograms of REDs. Right: KS statistic $D$. The KS statistic amplifies the differences in the time domain in cases of attacks (rows 2 and 3), while remaining small in the case of normal behavior (row 1).}\label{fig:attack12}
\vspace{-10pt}
\end{figure}

A simple way to detect anomaly is using a hard threshold on each individual error point, based on Gaussian assumption \cite{malhotra2015long} mean + 3 * std on the validation set. However, the Gaussian assumption is not true, because the reconstruction errors are non-negative and ``long-tail'' (Figure \ref{fig:attack12}). Besides, the threshold value significantly affects the accuracy and relies highly on the selection of the validation set, making it not stable nor reliable.




Hence, we enhance the detection by using the assumption-free Kolmogorov\text{-}Smirnov (KS) test to determine if the observed RED is the same distribution as the reference RED, without Gaussian assumptions. The KS test is a nonparametric test of one-dimensional probability distributions. It can be used to distinguish if two sets of samples are from the same distribution. The Kolmogorov\text{-}Smirnov statistic for two sets with $n$ and $m$ samples is:
\begin{align}\label{equ:KSstatistic}
D_{n,m} = sup_{x} |F_{n}(x)-F_{m}(x)|
\end{align}

where $F_{n}$ and $F_{m}$ are the empirical distribution functions of two sets of samples respectively, i.e. $F_{n}(t)=\frac{1}{n}\sum_{i=1}^{n}1_{x_i\leq t}$, and $sup$ is the supremum function. The null hypothesis that the two sets of samples are i.i.d. sampled from the same distribution, is rejected at level $\alpha$ if:
\begin{align}\label{equ:KStestinequ}
D_{n,m}>c(\alpha)\sqrt{\frac{n+m}{nm}}
\end{align}
where $c(\alpha)$ is a pre-calculated value and can be found in the standard KS test lookup table.

Suppose the reference and questioned RED are $D_1$ and $D_2$. We draw $m$ and $n$ independent samples from the two distributions, respectively. We calculate the KS test statistic, i.e., $D_{n,m}$ in Eq \ref{equ:KSstatistic}. Then we define a significance level $\alpha$, i.e. the probability of detecting a difference under the null hypothesis that samples are drawn from the same distribution. We reject the null hypothesis, i.e., recognize the samples as anomalies, if the inequality in Eq (\ref{equ:KStestinequ}) holds. 

We show that our enhanced deep learning approach is effective in amplifying small changes of control logic code in the PLC to large KS statistics $D$. We show three representative examples, Testing Normal (row 1), Attack 1 (row 2) and Attack 2 (row 3), in Figure \ref{fig:attack12}. \ding{172} The left column shows the reconstruction errors in the time domain. We observe that, in general, the reconstruction errors of Attacks (red in the 2nd and 3rd row) are slightly larger than the testing normal scenario (green in the 1st row). \ding{173} The middle column shows the histograms of the REDs. We observe that the RED of testing normals (green, row 1 middle) is more similar to the reference RED (blue, row 1 middle), than the RED of attacks (red, rows 2 and 3). \ding{174} The right column shows the KS test statistics $D$. We find that $D$ is significantly larger under the attacks (rows 2 and 3) than testing normal (row 1). This gives us assurance that our enhanced deep learning method (LSTM + RED) can indeed detect attacks which materialize as only small code changes in the PLC in power-grid systems. We show the quantitative comparison results in Section \ref{sec:Evaluation:accuracy}.

\section{Evaluation}\label{sec:Evaluation}
\subsection{System Configuration}\label{sec:implementation:config}

We implement a prototype of our proposed detection system. Our target system is a real power-grid controller sub-system, running the real deployed power-grid controlling code. The control codes are written in Structured Text (IEC 61131-3 standard), and run on the PLC as a multi-threaded program (e.g., 23 threads as I/O, controls, etc.). The 23 threads in the Wago PLC process being monitored correspond to the main thread and 22 threads. When the PLC powers on, 7 threads are created. When an application is loaded, another 15 threads will be killed and re-spawned with a different thread ID (TID). The last thread (PLC\_Task) is the one specifically running the loaded application (utilizing the other threads to do various tasks). We list the threads and their start phase in Table \ref{tab:threads}.

\begin{table}[h]
\centering
\caption{Threads running on Programmable Logic Controller}\label{tab:threads}
\scriptsize
\begin{tabular}{| C{0.2\linewidth}| C{0.7\linewidth} |}
\hline
  Start Time       & Thread(s) \\ \hline
  PLC powers on (7) & spi1, codesys3, com\_DBUS\_worker, 0ms\_Watch\_Thread, CAAEventTast, SchedExeption, Schedule \\ \hline
  Application loaded (15) & WagoAsyncRtHigh, WagoAsyncRtMed, WagoAsyncRtLow, WagoAsyncHigh, WagoAsyncMed, WagoAsyncLow, WagoAsyncBusCyc, WagoAsyncBusEvt, WagoAsyncBusEvt, WagoAsyncBusEvt, WagoAsyncBusEvt, ProcessorLoadWa, KBUS\_CYCLE\_TASK, ModbusSlaveTCP, PLC\_Task \\ \hline
  Main thread (1) & Main \\ \hline
\end{tabular}
\end{table}

The HPCs are monitored on a real PLC with an ARM Cortex A8 processor and a custom real-time Linux distribution. Data from 4 HPCs was collected for each thread: \ding{172} the number of cycles and \ding{173} the number of instructions executed, to provide the overall running status of the thread. \ding{174} The number of L1 cache misses, to reflect the locality and temporal properties of data usage. \ding{175} The number of branch instructions, to show the changes in the control flow of the threads. The total number of HPCs monitored is 4N if there are N threads. The HPCs are sampled at 1 kHz for each of the 23 threads. We sample 300,000 samples for each attack running. The data is relatively stable during our collection.

\subsection{Attack coverage}

We evaluate the system with six representative real attacks in Table \ref{tab:attacks}, which are known to cause damage to the power grid. The evaluated attacks are comprehensive enough to represent unknown attacks because we consider attacks against different components of PLCs, i.e. inputs (attack1), control flow (attacks 2,5), outputs (attack 4,6) and disabling the whole PLC (attack 3). The attacks include simulating the publicly reported malicious functionalities of Stuxnet \cite{langner2011stuxnet} and BlackEnergy \cite{2015powergridattack}, i.e., halting the controller and replaying prerecorded inputs to the legitimate code.

\begin{table*}[h]
\centering
\caption{Baseline and attacks against a power-grid PLC}\label{tab:attacks}
\resizebox{\textwidth}{!}{
\begin{tabular}{| c | p{0.65\linewidth} | p{0.15\linewidth} |}
\hline
    & \textbf{Description} & \textbf{Attack Type}\\ \hline
Baseline (Testing Normal)  & Baseline (a Proportional Integral Derivative (PID) controller) & None \\ \hline
Attack 1  & Overwrite the input (an additional line of code to overwrite the value of the input) & Input hijack \\ \hline
Attack 2  & Saturate the input (2 additional lines of code with IF condition on input value) & Control flow hijack \\ \hline
Attack 3  & Disable the PID control code (the entire PID block is commented out) & Entire Code hijack \\ \hline
Attack 4  & PID in "manual" mode (output is fixed) & Output hijack \\ \hline
Attack 5  & 2 PIDs Cascaded (input of a PID controller is sent through a second PID controller) & Control flow hijack \\ \hline
Attack 6  & Overwrite the output (an additional line of code to overwrite the value of the output) & Output hijack \\ \hline
\end{tabular}
}
\end{table*}

\subsection{Detection Evaluation} \label{sec:Evaluation:accuracy}

We use 6 performance metrics: accuracy, false positive rate, false negative rate, precision, recall and F1 score. \textit{Precision} measures the ratio of true positives among all predicted positives. \textit{Recall} measures the ratio of detected positives among all real positives. \textit{F1 score} is the harmonic mean of Precision and Recall, which balances these two metrics.

\bheading{Anomaly Detection Comparison.} We show the detection results of our proposed approach, and compare with conventional anomaly detection algorithms in Table \ref{tab:anomalybmark}. For fair comparison, we set the accumulated length of gathering RED (250) as the window size of conventional approaches. Note that all conventional approaches require heuristic features, while our approach takes raw data. We fine-tune the hyper-parameters and report the highest F1 score for the conventional methods.

Our proposed LSTM+RED detection methods (last row) achieve as high as 99.97\% accuracy and 0.9997 F1 score with no false negatives (all anomalies are detected). Our proposed LSTM+RED approach performs better than all evaluated methods, because conventional methods are not able to automatically extract inherent but complex features which can be used to model the normal behaviors of the target power-grid controller. In addition, our proposed RED provides a more robust way to detect anomalies by comparing distributions.


\begin{table*}[h]
\centering
\caption{Anomaly detection methods on detecting zero-day attacks against a real ARM Cortex A8 PLC controlling a real power-grid. Our proposed LSTM+RED approach performs better than all evaluated methods.}\label{tab:anomalybmark}
\resizebox{\textwidth}{!}{
{\begin{tabular}{| c | c | c |c |c |c |c |c |}
\hline

\textbf{Anomaly Detection Method} & \textbf{Key Hyper-parameters}         & \textbf{Accuracy}    & \textbf{False Positive}  & \textbf{False Negative}  & \textbf{Precision}   & \textbf{Recall}  & \textbf{F1} \\ \hline
One-Class SVM   & RBF Kernel, $\gamma = \frac{1}{N_{features}}$
& 0.849     & 0.104       & 0.621       & 0.267     & 0.379   & 0.313 \\ \hline
KNN                          & $N_{neighbors}=5$
& 0.930     & 0.000       & 0.766       & 1.000     & 0.235   & 0.380 \\ \hline
PCA                          & $N_{components}=10$
& 0.841     & 0.161       & 0.141       & 0.347     & 0.859   & 0.495 \\ \hline
Exponential Moving Average   & Smoothing Factor = 0.2, Smooth Window = 20
& 0.911     & 0.000       & 0.982       & 0.994     & 0.018   & 0.127 \\ \hline
Elliptic Envelope Estimation \cite{rousseeuw1999fast}  & Support Fraction = 1.0
& 0.867     & 0.146       & 0.000       & 0.406     & 1.000   & 0.578 \\ \hline
Robust Elliptic Envelope Estimation \cite{simpson1997introduction} &Support Fraction = $\frac{N_{samples} + N_{features} + 1}{2}$& 0.928     & 0.079       & 0.000       & 0.559     & 1.000   & 0.717 \\ \hline
Local Outlier Factor \cite{breunig2000lof}  & $N_{neighbors}=20$
& 0.929     & 0.000       & 0.783       & 1.000     & 0.218   & 0.357 \\ \hline
Isolation Forrest \cite{liu2008isolation}  & $N_{trees}=100$
& 0.949     & 0.056       & 0.000       & 0.640     & 1.000   & 0.780 \\ \hline
Bitmap Encoding \cite{wei2005assumption} & Sections = 4
& 0.911     & 0.001       & 0.968       & 0.799     & 0.032   & 0.061 \\ \hline
HBOS \cite{goldstein2012histogram} & $N_{bins}=10$, regularizer $\alpha=0.1$
& 0.899     & 0.111       & 0.001       & 0.473     & 0.999   & 0.642 \\ \hline
ABOS \cite{kriegel2008angle}  & $N_{neighbors}=5$
& 0.929     & 0.000       & 0.784       & 1.000     & 0.216   & 0.355 \\ \hline

\textbf{CRBM + RED (Explored)} &  $N_{hidden}=50$
& \textbf{0.957}      & \textbf{0.062}   & \
\textbf{0.023}    & \textbf{0.934}    & \textbf{0.977}    & \textbf{0.958}\\ \hline
\textbf{LSTM + RED (Proposed)} &  $N_{hidden}=256$
& \textbf{0.9997}     & \textbf{0.0005}  & \textbf{0.000}    & \textbf{0.9995}   & \textbf{1.000}    &\textbf{0.9997} \\ \hline
\end{tabular}}
}
\end{table*}

\begin{figure}
 \centering
  \includegraphics[width=\linewidth]{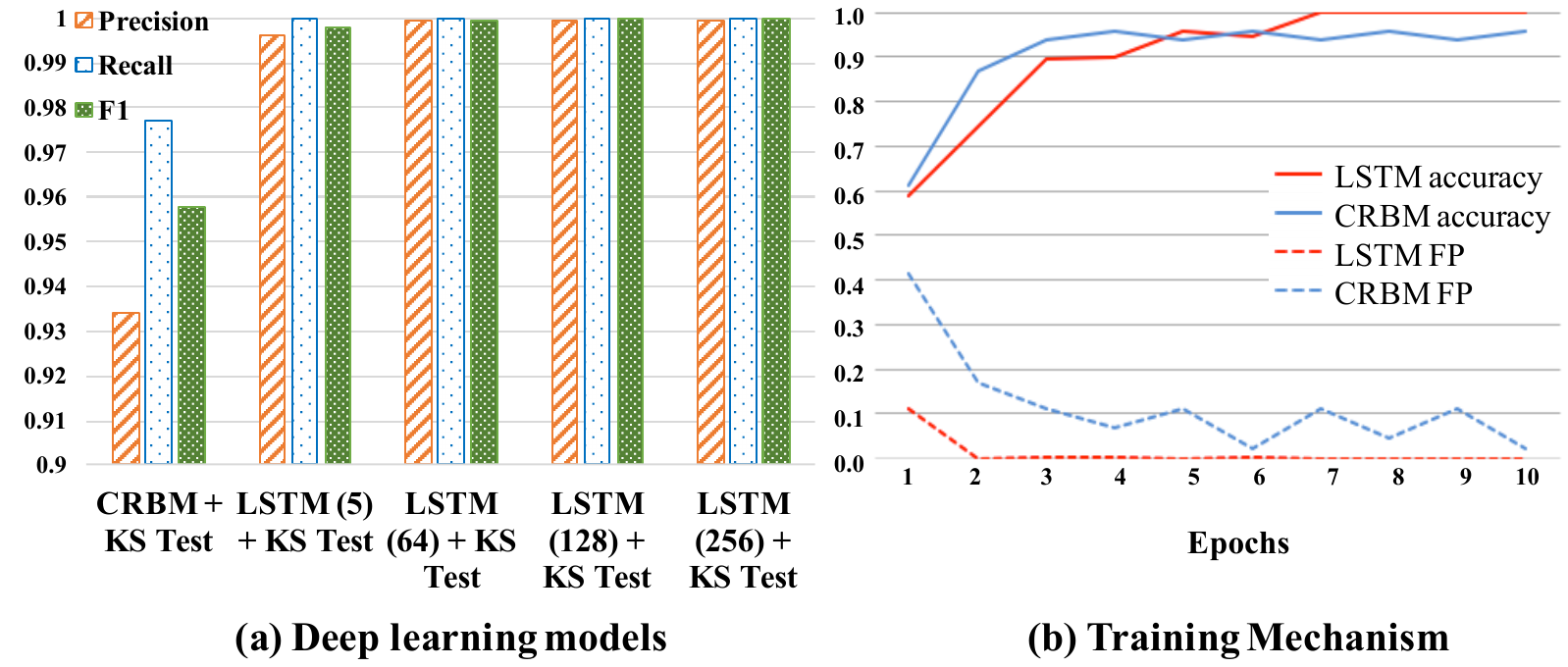}
  \caption{(a) Detection results of using LSTM and CRBM as the predicting module. LSTM performs better than CRBM. (b) Training mechanism for LSTM and CRBM. LSTM converges faster and achieves higher ultimate accuracy than CRBM.}\label{fig:models}
\end{figure}

\bheading{Temporal Learning Models Comparison.} We investigate how the types and architectures of the deep learning models affect the detection results. We use LSTM with different hidden nodes and an alternative CRBM model as the predictor in the experiment. In Figure \ref{fig:models} (a), we observe that the LSTMs perform better than the CRBM, because the LSTM automatically adjusts the ``memory window size'' while CRBM uses a fixed size. In Figure \ref{fig:models} (b), We find that LSTM convergences faster than CRBM, and achieves better final detection results. LSTM also becomes stable at the end of the training, however, CRBM continues to oscillate as we continue to train it.

By comparing different LSTM architectures in Table \ref{tab:ablation}, we observe that an LSTM with a medium number (128) of nodes in the hidden layer performs better than an LSTM with a small (5) or large (256) number of nodes in the hidden layer. Too few nodes in the hidden layer significantly limit the capability to represent the controller behavior, while too many nodes in the hidden layer cause over-fitting.

\bheading{Ablation Experiment of RED.} We evaluate the effectiveness and necessity of our proposed RED. In Table \ref{tab:ablation}, we compare the detection results with and without RED. The first 4 rows show the detection results of different LSTMs using hard thresholds, while the next 4 rows show the corresponding LSTM + RED for comparison. We observe that the RED improves all evaluated LSTM models and it provides insensitivity to the deep learning architectures (last 4 rows).

\begin{table}[h]
\begin{centering}
\caption{Effectiveness of enhancing DL with KS test}\label{tab:ablation}
\resizebox{\linewidth}{!}{
\begin{tabular}{|c|c|c|c|c|c|c|c|}
\hline
& Accuracy & False Positive & False Negative & Precision & Recall & F1     \\ \hline
LSTM(5)                   & 0.935    & 0.028          & 0.103         & 0.970     & 0.897  & 0.932  \\ \hline
LSTM(128)                 & 0.977    & 0.028          & 0.017         & 0.972     & 0.983  & 0.977  \\ \hline
LSTM(256)                 & 0.976    & 0.041          & 0.007         & 0.960     & 0.993  & 0.976  \\ \hline
LSTM(512)                 & 0.976    & 0.041          & 0.007         & 0.960     & 0.993  & 0.976  \\ \hline
LSTM(5) + RED              & 0.998    & 0.004          & 0.000         & 0.996     & 1.000  & 0.999  \\ \hline
LSTM(128) + RED            & 0.9998   & 0.0004         & 0.000         & 0.9996    & 1.000  & 1.000  \\ \hline
LSTM(256) + RED            & 0.9997   & 0.0005         & 0.000         & 0.9995    & 1.000  & 1.000  \\ \hline
LSTM(512) + RED            & 0.9997   & 0.0005         & 0.000         & 0.9995    & 1.000  & 1.000  \\ \hline
\end{tabular}
}
\end{centering}
\end{table}

\subsection{Latency Evaluation}


We list the components that contribute to the overall latency:

\bheading{\ding{172} LSTM Prediction Time $t_{predict}$.} This is the inference time for the LSTM model to predict the controller behavior, in terms of one HPC reading.

\bheading{\ding{173} Sampling Interval $t_{sample}$.} This is the interval between two HPC readings.

\bheading{\ding{174} Error Accumulation (EA) Time $t_{EA}$.} This is the time to gather one point of the reconstruction error. Note that in Eq (\ref{equ:reconstructionerror}), one point of the reconstruction error is accumulated among $L$ time frames. Therefore, $t_{EA}=max\{t_{predict}, t_{sample}\}\times L$.

\bheading{\ding{175} RED Collection Time $t_{RED}$.} This is the time to collect the testing reconstruction error distribution. This distribution is represented by $N_{EA}$ independent reconstruction errors in \ding{174}.

\bheading{\ding{176} KS-Test Time $t_{KS}$.} This is the time to compute the KS statistic and determine if an anomaly has occurred.


The detection latency $t_{detection}$, i.e. the time interval between the occurrence of the attack and the detection of the anomaly:
\vspace{-0.1cm}
\begin{align}
t_{detection}&= max\{t_{predict}, t_{sample}\}\times L\times N_{EA} + t_{KS} \label{eq:tD}
\end{align}
\vspace{-0.5cm}

We observe that Eq (\ref{eq:tD}) is dominated by the first term. Thus we look for the minimal $L \times N_{EA}$ that maintains false positives rate (FPR) $<10^{-4}$. We observe a significant FPR increase ($\sim$10.4\%) after $L \times N_{EA}$ = 250, thus we choose $L=10$ and $N_{EA}=25$ in our implementation.


\begin{table}[h]
\centering
\caption{Execution time for model training and inferencing, and KS testing, for detecting controller anomalous behavior.} \label{tab:latency} 
\resizebox{\linewidth}{!}{
\begin{tabular}{| c | c | c | c | c | c | c | c | c |}
\hline
     & \textbf{Training}  & $t_{network}$ & $t_{sample}$ & $t_{predict}$ & $t_{EA}$     & $t_{RED}$ & $t_{KS}$         & \textbf{Detection} \\ \hline
\textbf{GPU}  & \textbf{21.1 min}   & 100 ms   & 1 ms         & 0.53 ms       & 25 ms        & 250 ms    &   8.6 ms ${}^\star$ & \textbf{358.6 ms}        \\ \hline
\end{tabular}
}
\begin{tablenotes}
    \scriptsize
    \item ${}^{\star}$ KS test is implemented on the CPU.
\end{tablenotes}
\end{table}

Table \ref{tab:latency} shows the latency evaluated in our system. The total detection latency (386.5ms) is much shorter than human response and reaction time, i.e. a few seconds. Thus, our proposed method significantly reduces the response time and provides a quick mitigation mechanism.

\subsection{Security Discussion} \label{sec:solution:security}

All the evaluated attacks were successfully detected. We discuss further potential attack strategies and how they are defended by our proposed solution.

\bheading{Case 1: The attacker hides and waits for a specific time to launch an attack.} Our proposed method provides continuous and holistic protection of the controller. If the attack code keeps silent, it does not do any harm to the system. When the attack becomes active and influences the behavior of the controller, our experiments show that we can quickly detect it and reduce the damage to the system.

\bheading{Case 2: The attacker tampers with the computation.} It is harder for the attacker to tamper with the prediction and KS test because the deep learning prediction and KS test are not computed by software on the controller, but done in a protected module in our design.

\bheading{Case 3: Adversarial examples \cite{goodfellow2014explaining} against the learning model.} Generating adversarial examples against our system is generally hard, because the RED is generated from HPCs which are hard to mimic. 



\section{Related and Future Work}\label{sec:relatedwork}

Previous work has shown the feasibility of using HPCs for detecting malware \cite{ozsoy2016hardware, demme2013feasibility}, firmware-modification \cite{wang2015confirm} and kernel root-kits \cite{wang2013numchecker}. To the best of our knowledge, we are the first to use HPCs to detect anomalies in the critical power-grid controllers. Furthermore, we are the first to propose reconstruction error distribution of HPCs as a profile of normal behavior, while previous work only use conventional machine learning and hard thresholds. Previous work on using deep learning for improving security \cite{tang2016deep, pascanu2015malware, he2017machine} and protecting deep learning systems \cite{he2019sensitive, zhang2018privacy} have been proposed.

We detect anomalous behavior in this paper, which may be attacks or benign. Further situation dependent testing needs to be done to determine if it is a known or zero-day attack -- a topic for future work. Our proposed defense cannot detect an attack when the HPCs are not available, e.g. the PLC is shut down, the PLC does not provide HPC readings, or the HPCs reading process is terminated by a malicous process. 

\section{Conclusion}\label{sec:conclusion}

We propose a new data-driven approach to detect anomalous behavior in power-grid controllers. Our method leverages a temporal deep learning model, and is enhanced by our proposed reconstruction error distribution (RED) of low-cost hardware performance counters (HPCs). A key advantage of our approach is that only normal data is used, and no prior-knowledge about the attack data is needed, thus enabling detection of zero-day attacks. We evaluate our detection system on a real programmable logic controller used in power-grid substations. We show the significant improvement of using enhanced deep learning compared to 11 conventional anomaly detection approaches. Experiments show that our proposed system achieves nearly zero false positives and low latency.

\bheading{Acknowledgment.} We thank Prashanth Krishnamurthy, Farshad Khorrami and Ramesh Karri from NYU for giving us access to the raw data on the baseline behavior of the PLC.

\bibliographystyle{plain}
\bibliography{main}

\end{document}